# Loss of vimentin intermediate filaments decreases peri-nuclear stiffness and enhances cell motility through confined spaces

Alison E. Patteson[1,2]*, Katarzyna Pogoda[1,3], Fitzroy J. Byfield[1], Elisabeth E. Charrier[1], Peter A. Galie[1,4], Piotr Deptuła[5], Robert Bucki[5], and Paul A. Janmey[1]*

1 Institute for Medicine and Engineering, University of Pennsylvania, Philadelphia, PA 19104
2 Physics Department, Syracuse University, Syracuse, NY 13244
3 Institute of Nuclear Physics, Polish Academy of Sciences, PL-31342 Krakow, Poland
4 Department of Biomedical Engineering, Rowan University, Glassboro, NJ 08028
5 Department of Microbiological and Nanobiomedical Engineering, Medical University of Białystok, Mickiewicza 2C, Białystok, Poland

*Corresponding author. Email: aepattes@syr.edu (A.E.P.); janmey@pennmedicine.upenn.edu, (P.A.J.)

The migration of cells through tight constricting spaces or along fibrous tracks in tissues is important for biological processes, such as embryogenesis, wound healing, and cancer metastasis, and depends on the mechanical properties of the cytoskeleton. Migratory cells often express and upregulate the intermediate filament protein vimentin. The viscoelasticity of vimentin networks in shear deformation has been documented, but its role in motility is largely unexplored. Here, we studied the effects of vimentin on cell motility and stiffness using mouse embryo fibroblasts derived from wild-type and vimentin-null mice. We find that loss of vimentin increases motility through small pores and along thin capillaries. Atomic force microscopy measurements reveal that the presence of vimentin enhances the perinuclear stiffness of the cell, to an extent that depends on surface ligand presentation and therefore signaling from extracellular matrix receptors. Together, our results indicate that vimentin hinders three-dimensional motility by providing mechanical resistance against large strains and may thereby protect the structural integrity of cells.

**Introduction**
Many important biological processes, such as embryogenesis [1,2] and wound healing [3,4], depend on the ability of cells to move through the tight constricting spaces in tissues. This same ability - under disruptive conditions - can lead to invasive cell migration and cancer metastasis [5,6]. A key event that triggers cell migration is the epithelial-mesenchymal transition (EMT), in which non-migratory epithelial cells lose cell-cell adhesions and transition to migratory polarized mesenchymal cells [1,2]. During this transition, cells express and upregulate the intermediate filament protein vimentin (Fig. 1a), a wide-spread marker of EMT and mesenchymal cells [1,7]. Vimentin's role in cell motility has been implicated in aggressive tumors throughout the body, including the prostate, breast, and lung, and high levels of vimentin expression in cancer cells correlate with accelerated tumor growth and poor prognosis [8].

*In vivo*, cells move by either degrading the surrounding extracellular matrix (ECM) or deforming their bodies in order to squeeze through narrow openings of the ECM or conform to highly curved fibrous tracks such as nerves or small blood vessels [9,10]. The interstitial spaces of tissues are complex environments that contain pores, 0.1 to 30 μm in diameter [11,12], and pre-existing tracks, 100-600 μm in length [13], that act as paths for neutrophil [9] and cancer cell [9,12] migration. For small pores (2-3 μm), nuclear size and rigidity significantly limit migration [9,14,15], and when cells do squeeze through small pores, the large strain imposed on the cell and nucleus can lead to nuclear membrane rupture [16,17], accumulated DNA damage [16-18], and enhanced genome variation [18].

A robust feature of vimentin intermediate filaments is their formation of a cage-like network that encircles the cell nucleus (Fig. 1a) [19] and contributes to the mechanical integrity of the cell [20-22]. Even under conditions in which the peripheral vimentin networks is destabilized, such as growth on soft substrates, the perinuclear vimentin cage remains intact [23]. Reconstituted vimentin networks are viscoelastic materials that significantly stiffen at large strains and are capable of withstanding extreme deformations that cause other cytoskeletal polymers (actin and microtubule) to fail [24,25]. Vimentin's unique mechanical properties are hypothesized to confer enhanced resistance to cellular deformations especially at large strains [22,24]. However, experimental demonstrations remain sparse, and little is known about how vimentin affects cell motility through constricting spaces.

**Results**
**Loss of vimentin increases cell motility through constricting spaces**
To model cell migration in 3D environments, we designed micro-fluidic devices with confining channels. The channels were large enough to allow the nucleus to pass through yet small enough to constrict the vimentin perinuclear cage (Fig. 1, **SI Fig 1-3**). We observed that both wild-type (vim +/+) and vimentin-null (vim -/-) mEF exhibited spontaneous migration into collagen I-coated micro-channels (Fig. 1b) at migration rates similar to *in vivo* speeds (50-300 um/hr) [26] observed for amoeboid cancer cells. These persistent motions occurred in the absence of both chemical gradients and applied pressure gradients across the channels, consistent with previous studies using 3T3 fibroblasts [27].

Here, we found that the ability to pass through the constrictions differs strikingly for cells with and without vimentin. Surprisingly, the loss of vimentin enhanced migration, enabling cells to enter the channels faster and increasing the probability of crossing the channel (Fig. 1b iii, **SI Movie 1 & 2**). Cells that do not pass through either change direction or get stuck, events that occur more frequently in wild-type than vimentin-null mEF (*c.f.* Fig. 2, **SI Movie 3**). Similar behaviors were observed in 3T3 fibroblasts, which also showed a marked decrease in motility compared to fibroblasts lacking vimentin (**SI Fig. 4**).

To test the effects of vimentin on motility in fibrillar biopolymer matrices, we embedded the wild-type and vim -/- mEF in collagen I gels and tracked their motion over 17 hr (Methods). Bright-field images of the cells in the gel and sample trajectories are shown in Fig. 1c. In the collagen gels, the cells are highly confined (expected average 1-2 μm pore size [28]), and their motility is limited. Under these conditions, the vimentin-null cell behavior is more varied than that of wild-type mEFs, yet a subset of vim -/- mEFs squeeze through the narrow pores and travel much greater distances than wild-type mEFs.

The micro-fluidic channels and collagen gel experiments highlight an unexpected trend: vimentin hinders motility through 3D confining spaces. This result contrasts with previous studies that show vimentin enhances motility on unconfined rigid substrates [20,29]. What effect does cell confinement have on vimentin's role in motility? To address this question, we compared the instantaneous speeds of cells measured in environments with varying degrees of confinement, which include unconfined 2D glass substrates, micro-channels of various width, and highly-confined collagen gels (Fig. 1d, Methods). On the glass slides, the instantaneous speeds are statistically similar (Fig. 1d), although wild-type mEF motion is more persistent than vim -/- (**SI Fig. 5**). In the micro-channels, cells experience increasing levels of confinement as the width decreases from 20 to 10 µm. For the vimentin-null cells, the confinement increases cell speed by a factor of 2.25 ($p < 0.001$), whereas wild-type speed increases by only 24% ($p = 0.058$). In the porous gels, cell speeds are significantly lower compared to the larger micro-channels. Unlike the glass slides, however, vimentin decreases cell speed in the 3D gels that confine the cells.

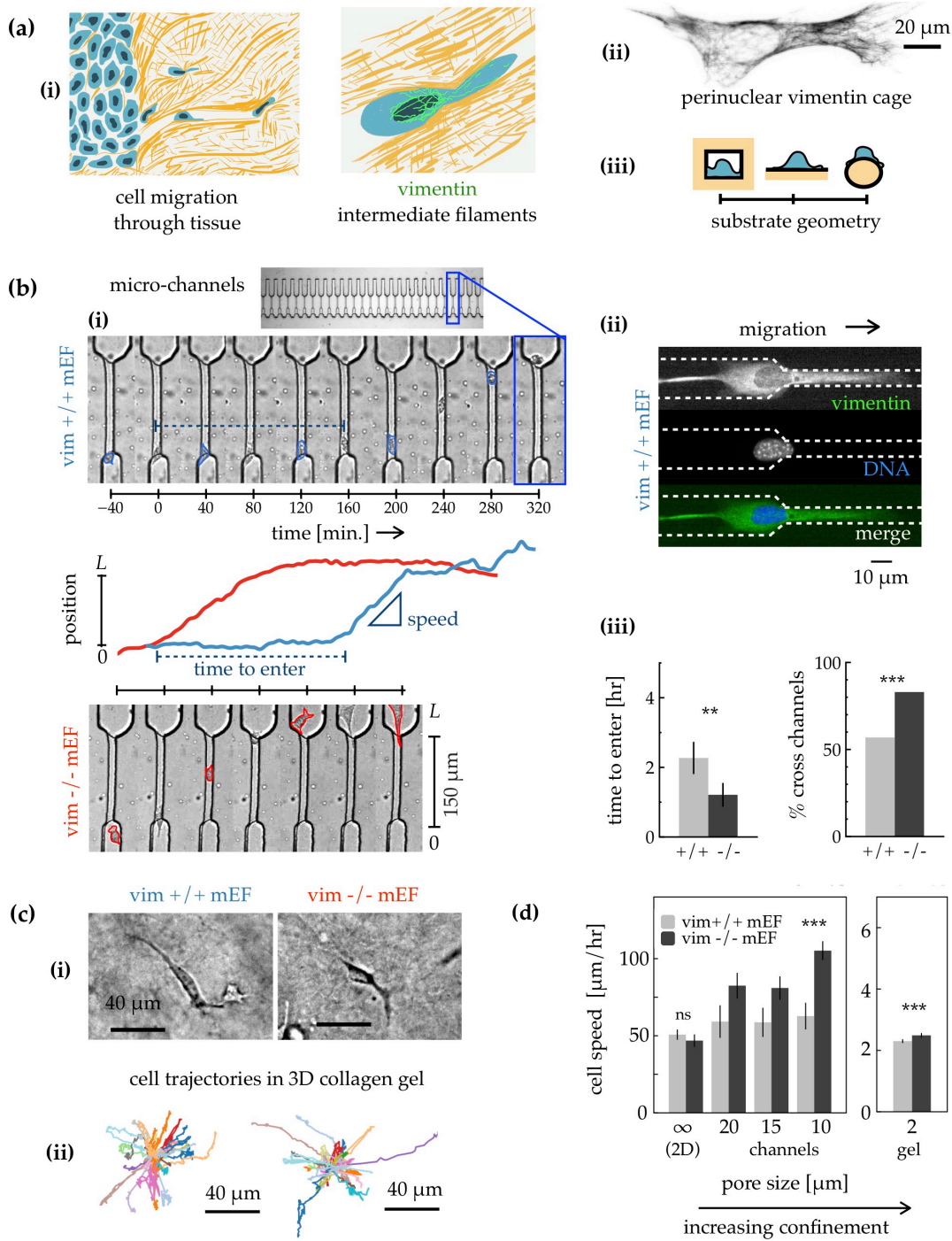

**Figure 1: Vimentin decreases cell motility through constricting spaces (a)** (i) Migratory cells often express the intermediate filament vimentin during important events such as the epithelial-mesenchymal transition, wound healing, and aggressive tumor growth. Vimentin forms a peri-nuclear cage, which could hinder migration through constricting interstitial spaces. (ii) Immunofluorescence image showing vimentin in a wild-type mouse embryo fibroblast (mEF). (iii) Unlike the rigid flat surfaces common to cell culture methods, interstitial spaces are three-dimensional environments containing small pores and highly curved tracks, such as along nerve fibers. **(b)** Cells are seeded in PDMS mico-fluidic devices in the absence of applied pressure and chemical gradients across the channels. (i) Brightfield images showing wild type (vim +/+) and vimentin null (vim -/-) mEF migrating through collagen I-coated micro-channels (SI Movies 1&2). Cells circled for visualization. (ii) Immunofluorescence image showing co-localization of vimentin and nucleus for wild-type mEF entering channel (Fibronectin-coated, 12 hr). (iii) Vim +/+ mEF take more time entering and are less likely to pass through micro-channels (collagen-coated) than vim -/- mEF. **(c)** (i) Bright field images of vim +/+ and vim -/- mEFs cultured in 3D collagen gels (24 hr, 2 mg/mL). (ii) Cell trajectories (over 17 hr) show that vim -/- mEF migrate more through gel than vim +/+ mEF. **(d)** Cell speed depends on vimentin and confinement. On unconfined glass slides, cell speed is similar between the two cell lines ($N = 45+$ cells each). In the micro-channels, the speed of vim -/- mEF increases with decreasing channel width at a significantly larger rate than vim +/+ mEF ($N = 35$-$60$ cells each). In collagen gels where confinement is strongest, vim -/- are faster than vim +/+ mEFs. Error bars denote standard error. Denotation ***, $p < 0.001$; **, $p < 0.01$.

Taken together, these experiments (Fig. 1) demonstrate that vimentin can significantly hinder migration and that this effect is stronger in more constricting environments. Increased cell speed with confinement has been observed previously [27,30] and is implicated with transitions to faster migration modes, such as amoeboid migration [30]. Moderate confinement induces changes in the cytoskeleton, such as an accumulation of actin in the cortex and an alignment of microtubules that is important for migration along 3D tracks [31,32]. In our micro-channels, application of microtubule and actin inhibitors causes a rapid decrease in motility for vim +/+ and vim -/- mEFs, indicating that the 3D motility observed here depends on actin and tubulin (**SI Fig. 6 & 7**). The results shown here (Fig. 1) also highlight that vimentin plays an important role in controlling 3D motility.

**3D fibroblast motility depends on vimentin and surface integrin-coating**
We observe differences in cell behavior between surfaces coated with collagen I and fibronectin (Fig. 2). For both cell lines, the projected spread area was greater in fibronectin-coated micro-channels compared to collagen I (Fig. 2a, **SI Fig. 8**), similarly to other stiff surfaces [22]. For both ligands, the average spread area was approximately the same between the two cell lines, suggesting similar cell volumes despite having different cell speeds (Fig. 2a). Cells lacking vimentin were faster than wild-type mEF for both coatings: yet showed a 22% decrease ($p = 0.011$) on fibronectin compared to collagen I; whereas normal cell speeds were constant.

Ligand coating impacted emergent behaviors in the micro-channels, such as twirling and persistency (Fig. 2b). When seeded on fibronectin, both cell types were more likely to exhibit twirling (**SI Movie 4**), appearing to move in three-dimensional helices, compared to their movement on collagen. Interestingly, twirling was more frequent in normal cells than cells without vimentin. In addition, persistent migration – characterized here by the percentage of cells that pass through the channel – depended on the surface ligand coating and vimentin. Wild-type mEF became more persistent in the presence of fibronectin compared to collagen, whereas vimentin-null mEF were persistent on both.

Cellular response to extracellular ligands is important for healthy tissue maintenance, for example fibronectin stimulates fibroblast migration in damaged tissues and collagen helps maintain homeostasis [33,34]. The different effects of collagen-I and fibronectin on wild-type mEF motility suggest distinct biochemical signals from different integrins that engage these ligands. Overall, the results in Fig. 2 indicate a role of vimentin in regulating cell migration in response to different ligands, which includes promoting persistent migration in the presence of fibronectin, a result that may be important for understanding wound healing models.

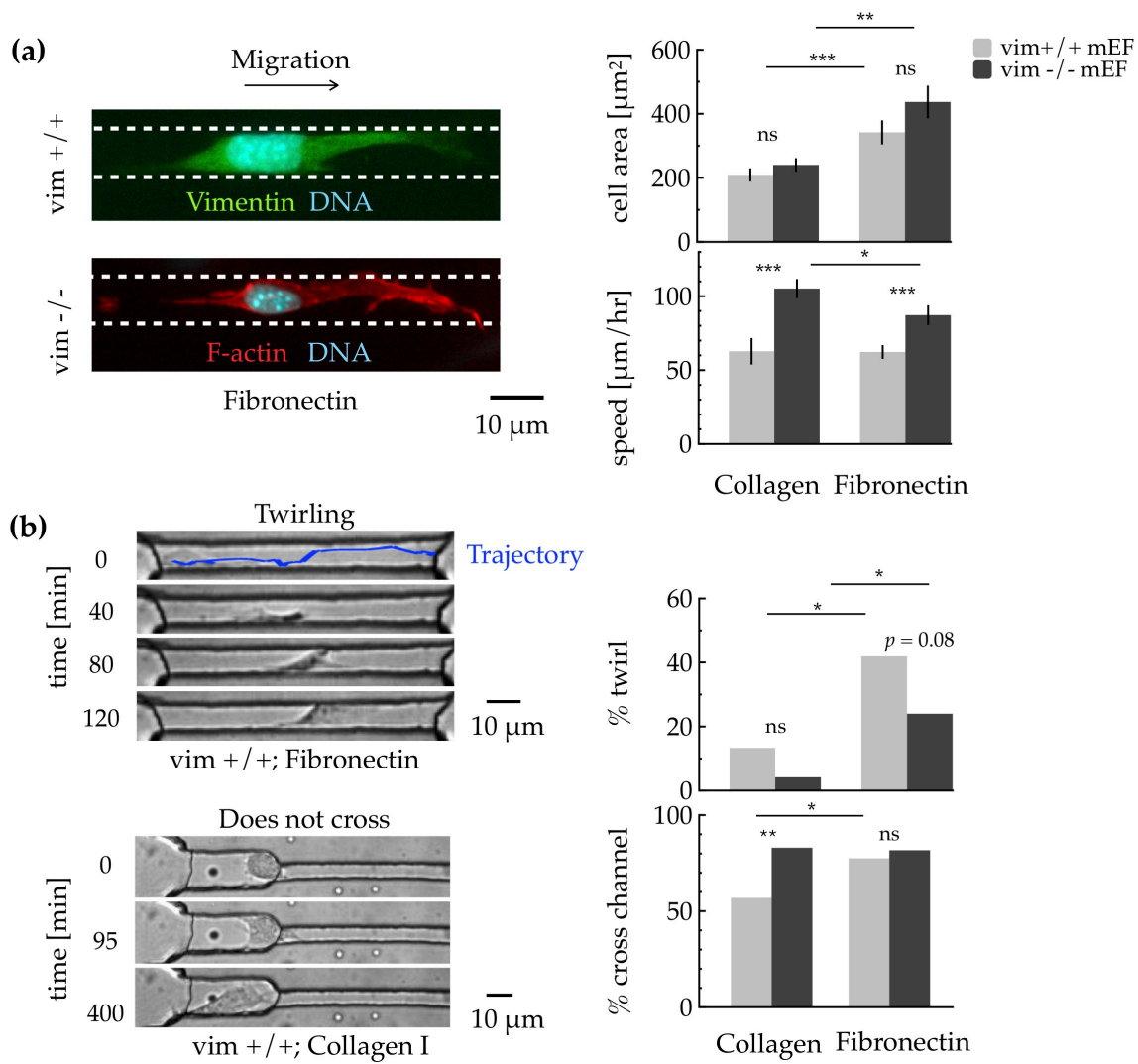

**Figure 2: Vimentin and surface-ligand coatings control 3D cell migration (a)** Left: Immunofluorescence images of vim +/+ and vim -/- mEF in micro-channels coated with fibronectin. Right: Both cell types are more spread in channels coated with fibronectin than collagen I. Cells lacking vimentin slow down on fibronectin compared to collagen-coated channels. **(b)** In the microchannels, cells display different behaviors, such as twirling, getting stuck, and switching directions (SI Movies 3&4). Twirling appears more often in vim +/+ than vim -/- mEF and is more frequent on fibronectin-coated channels than collagen. Normal mEFs are more likely to cross channel in the presence of fibronectin. Error bars denote standard error. Denotation ***, $p < 0.001$; **, $p < 0.01$; *, $p < 0.05$.

**Vimentin hinders migration along highly-curved capillaries**
Similar to the behavior in micro-channels, vim -/- cells migrate faster than vim +/+ when cultured on glass capillaries with diameters ranging from 10-20 µm (Fig. 3a). Migration of vim -/- cells was also sensitive to the ligand coating of the substrate, with a higher migratory speed observed on collagen-coated capillaries compared to those coated with fibronectin (Fig. 3b). Both cell lines tended to have a high aspect ratio and were elongated in the axial direction rather than the radial direction of the capillaries, possibly due to the high bending energy of F-actin [35] (Fig. 3c). Nuclei of vim +/+ cells, cultured on capillaries, were observed to be off-center relative to the distribution of vimentin within the cell (Fig. 3d).

**Vimentin stiffens cells on substrates coated with collagen but not fibronectin**
To understand vimentin's role in the ability of cells to move and deform, we used atomic force microscopy (AFM) experiments to measure the stiffness of normal and vimentin-null mEFs (Fig.

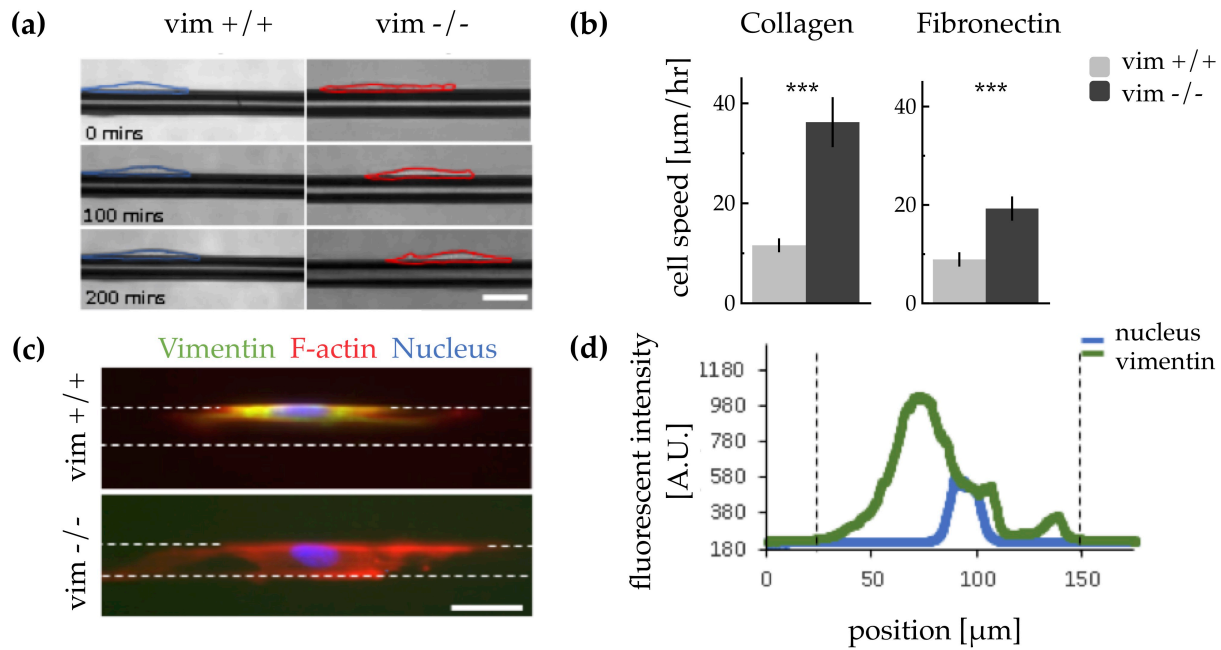

**Figure 3. Loss of vimentin increases cell migration on highly curved capillaries** (a) Bright field time-lapse images of vim +/+ and vim -/- mEF cultured on collagen-coated capillaries, taken over approximately 3 hr. (b) Vim -/- cells migrate faster than vim +/+ cells on both collagen- and fibronectin-coated capillaries (N=17-27 cells per condition). While the speed of vim +/+ cells is similar on both collagen- and fibronectin-coated substrates, the speed of vim -/- cells is higher on collagen. (c) Fluorescence images showing the typical distribution of vimentin and F-actin in a vim +/+ mEf and F-actin in a vim -/- mEF, cultured on collagen coated capillaries (21 hr). Dotted lines indicate the boundaries of the capillary. (d) Line profiles of a fluorescent image showing the typical position of the nucleus (stained with DAPI) relative to the distribution of vimentin (stained with anti-vimentin ab) in a vim +/+ cell on a collagen-coated capillary. Dotted lines indicate boundary of vimentin staining. A.U: Arbitrary units. Error bars denote standard error. Denotation ***, $p < 0.001$.

3). Notably, the loss of vimentin does not significantly change the expression levels of either F-actin and microtubules between these two cell lines [21], and we observe no obvious differences in the organization of these networks from fluorescence staining on collagen- (Fig. 3a) or fibronectin coated glass surfaces (**SI Fig. 9**). We measured cell stiffness in the perinuclear region of the cell, sampling multiple positions just outside the periphery of the nucleus, where filamentous vimentin is abundant (Fig. 3b). Cells were cultured on glass coverslips and then subjected to periodic indentation by an AFM tip, which was applied at a constant force amplitude of 3 nN and frequency 0.4 Hz.

We found that cell's perinuclear stiffness depended strongly on the presence of vimentin and the type of adhesive ligand used for coating. On collagen-coated surfaces, the stiffness of normal mEFs was approximately 13 kPa, which was 2.5x times greater than for vimentin-null mEFs ($p<0.001$). On fibronectin, the stiffness of normal mEFs was lower (7.5 kPa) and more equal to the stiffness of vim -/- mEFs (5 kPa, $p = 0.09$). These trends are striking compared to previous

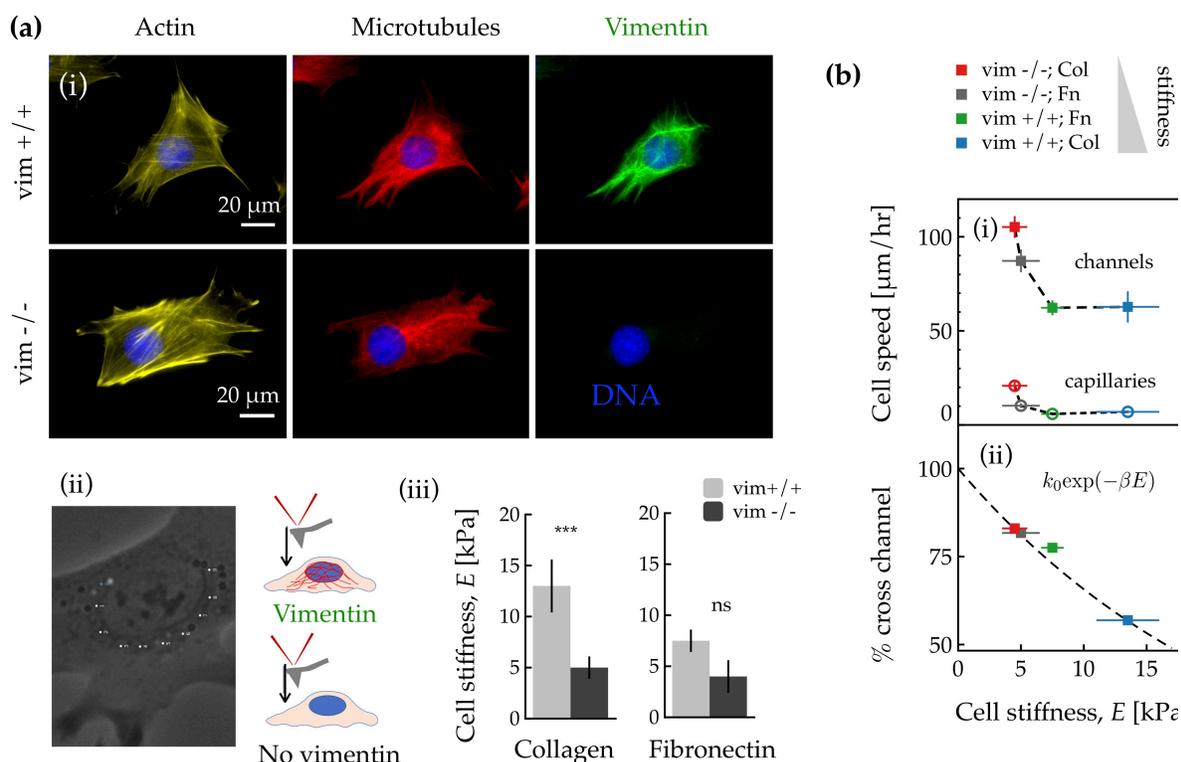

**Figure 4: Vimentin increases cell stiffness on collagen coated surfaces and impedes 3D migration (a)** (i) Immunofluorescence images of vim +/+ and vim -/- mEF showing the cytoskeletal networks: actin, microtubule, and vimentin (as shown on collagen-coated glass slides). Vim -/- mEFs maintain robust actin and microtubule networks. (ii) To measure peri-nuclear stiffness, an AFM tip was used to probe the region juxtapose to the nucleus, sampling at multiple positions (as labeled) to account for cytoskeletal variations. (iii) Perinuclear stiffness depends on the presence of vimentin and the type of adhesive ligand used for coating: vim +/+ mEFs are stiffer than vim -/- mEF on glass slides coated with collagen but not fibronectin. ($N = 30+$ cells per condition). **(b)** (i) For the glass capillaries and micro-channel experiments, cell speed decreases with increasing cell stiffness: dotted lines connect points to guide the eye. Ligand coating for each condition, from left to right, collagen (Col), fibronectin (Fn), Fn, Col. (ii) Percentage of cells crossing the micro-channels versus cell stiffness $E$. Dashed line is fit of the data to activated model $k_0 \exp(-\beta E)$, which suggests that stiffness limits migration by increasing the total work $\beta E$ for cells to cross channels. Error bars indicate standard error. Denotation ***, $p < 0.001$.

AFM experiments [22], which probed the endoplasmic region of these two cell lines, and found that vim +/+ mEFs were stiffer than vim -/- only when cells with maximal spread areas (spread area > 14, 000 µm$^2$) were compared. The experiments shown here (Fig. 3) highlight an important role of vimentin in mediating stiffness in the perinuclear region of the cell, a role that also depends on the surface ligand coating.

**Cell speed in 3D environments correlates with cell stiffness**
Consistent with the hypothesis that cytoskeletal stiffness hinders cell migration through constricted spaces, we find that 3D cell speeds monotonically decrease with cell stiffness. Here, cell stiffness varies depending on surface ligand coating and the presence of vimentin. A scatter plot of population-averaged cell speed versus cell stiffness indicates a general trend that cell stiffness suppresses cell speeds in 3D environments, as shown in Fig. 4b for both the glass capillaries and micro-channels. Interestingly, on the 2D glass coverslips, vim +/+ and vim -/- cell speeds do not depend on cell stiffness. Our results (Fig. 4bi) suggest that softer cells move faster than stiff ones in 3D and vimentin's contribution to stiffness thereby decreases migration.

The ability of cells to cross the micro-channels also decreases with cell stiffness (Fig. 4bii). To interpret this result, we suggest a minimal model that treats crossing the channel as an activated process with rates controlled by effective energy barriers needed to be overcome for cells to migrate across the channel. Assuming that peri-nuclear stiffness $E$ presents a significant energy barrier for the cell to cross the channel, the flux of cells is given by $k_0 \exp(-\beta E)$, where $k_0$ is the flux in the absence of cell deformation and $\beta$ is a constant determined by details of the coupling between cell deformation and channel geometry. As cell elasticity increases, the work required to cross the channel increases and the flux decreases by a factor $\exp(-\beta E)$, consistent with the observed decrease in cell flux (Fig. 4bii).

**Vimentin's contribution to peri-nuclear stiffness limits cell migration through small pores**
To test the limits of vimentin's effects on migration, we performed confined motility experiments using polycarbonate Transwell membranes with pore diameters varying from 3 to 8 µm, significantly smaller than the micro-channels. Cells were seeded on top of the membrane and allowed to adhere and migrate for 15 hr before being fixed and counted (Fig. 5a, Methods). The migration rates through the small pores depended on whether the membranes were coated with collagen or fibronectin (Fig. 5b). For collagen, vimentin-null mEF crossed at higher rates than wild-type mEF; even for the 3 µm pores – where the nucleus is expected to be limiting [9,14,15] – the presence of vimentin significantly impacted motility, decreasing speed by a factor of 4.2 ($p = 0.004$). For fibronectin, motility rates were equalized between the cell lines.

Variations in pore migration (Fig. 5b) may be attributed to differences in cell stiffness: to cross smaller pores, cells must undergo larger strains and expend more energy, which increases with cell stiffness. Here, we estimate that the cell compressional strain varies from 0.47 to 0.80 based on the geometry of the pores (Fig. 5c, Methods). To interpret how cell migration depends on cell stiffness and cell strain, we extend the activated model of constricted migration, assuming that the energy barrier to cross the pore is proportional to the cell stiffness $E$ and the parameter $\epsilon_0 \epsilon$, where $\epsilon$ is the compressional strain needed to cross the membrane and $\epsilon_0$ is a pre-strain based upon internal stresses that develop as cells adhere and spread on the surface. The flux of cells then decreases by a factor $\exp(-\beta\epsilon_0\epsilon E)$ for increasing cell stiffness $E$ (a function of cell type and ligand coating) and strain $\epsilon$ (based on pore size alone). By fixing cell stiffness $E$ to values from the AFM measurements (Fig. 4), we fit the four curves in Fig. 5c to $k_0\exp(-\beta\epsilon_0\epsilon E)$ and obtain a single value for the constants $\beta\epsilon_0 = 0.18$ kPa$^{-1}$ and $k_0 = 16\%$. The model seems to capture the main features of the experimental data and further supports the idea that constricted cell migration depends strongly on vimentin peri-nuclear stiffness.

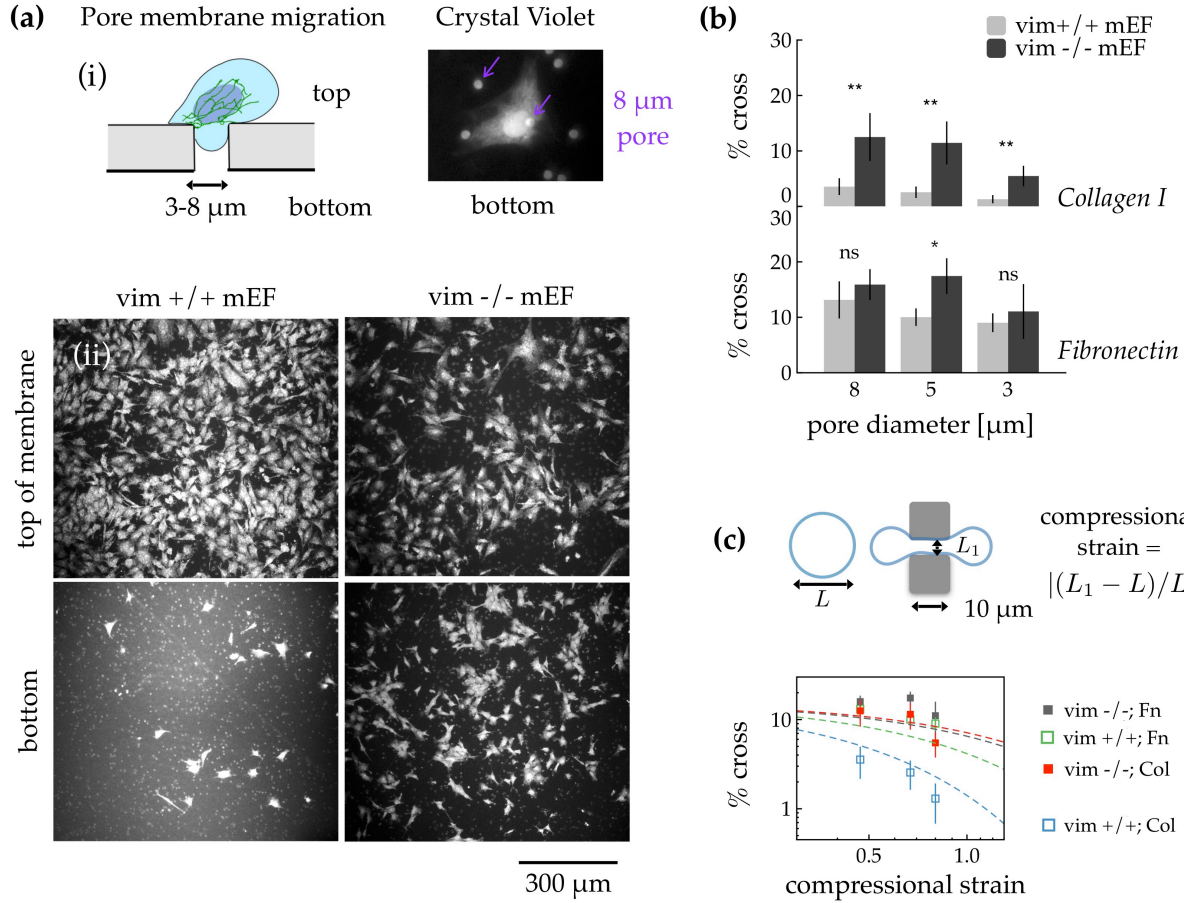

**Figure 5: Vimentin intermediate filaments impede migration through nucleus-limiting pores (a)** (i) The role of vimentin in constricted cell migration is probed using Transwell membranes with pore diameters ranging from 3 to 8 μm. Vim +/+ and vim -/- mEF are seeded on top of membranes and allowed to migrate for 15 hr, then fixed and stained with crystal violet. (ii) Fluorescence images of cells localized to the membrane top and bottom show that more vimentin-null mEF translocate pores than wild type mEF, as shown for 8 μm collagen-coated pores. **(b)** The percentage % of cells that cross the filters depends on pore size and surface ligand coating. Fibronectin equalizes the crossing rates of the two cell types. **(c)** Smaller pore sizes require higher cell strain and more work to passage through. The % cross versus the compressional strain is fit to the activated model $k_0\exp(-\beta\epsilon_0\epsilon E)$, where $E$ is the AFM-measured cell stiffness; the fitting parameters $k_0 = 16\%$ and $\beta\epsilon_0 = 0.18$ kPa$^{-1}$ are each a single constant across all four data sets. Error bars indicate standard error. Denotation **, $p < 0.01$; *, $p < 0.05$.

### Discussion

Intermediate filaments are important in development, tissue maintenance, metastasis, and disease. Using a knock-out model for vimentin, we have found that vimentin hinders cell motility through small pores and along thin capillaries, in contrast to previous studies, which have shown that vimentin enhances motility on 2d plastic cell culture substrates. The strong effect of vimentin on 3D motility may be surprising, given that intermediate filaments are much softer than F-actin or microtubules and thus easiest to deform to large strains. Our results here indicate that strain-stiffening and resistance to breaking [24] may be the important mechanical features through which vimentin contributes to migration. It is possible that vimentin's role in regulating actin stress fiber assembly through rho-A [36] and possible other signaling mechanisms contribute to differences in motility for cells with and without vimentin, although no difference in actin expression levels have been reported for these two cell lines [21].

The physiological effects of intermediate filaments in 3D cell migration remain unclear. Vimentin expression reflects a phenotypic characteristic that contributes to normal cell migration during EMT but also aggressive behavior of tumor cells. The use of vimentin-disrupting drugs such as withaferin A have been proposed as effective therapy for cancer treatment [8]. Recent studies have shown that modification of keratin intermediate filament organization in pancreatic cancer cells by the phospholipid sphingosylphosphorylcholine (SPC) enhanced cell deformability and migration in micro-channels [37], which is suggested to correlate with increased metastatic potential. Here, we find that the loss of vimentin in a knock-out mouse model similarly increases cell deformability and enhances constricted cell migration.

These results might seem counter-intuitive, considering that as epithelial cells undergo EMT and become more migratory they upregulate vimentin to replace cytokeratin rather than downregulating intermediate filaments generally. However, the greater localization of vimentin around the nucleus compared to other cytoskeletal filaments could help provide a more detailed understanding of why migratory cells express vimentin. One possible explanation is that the perinuclear vimentin network is required to cushion the nucleus or the cortical actin network during extreme strains associated with movement through constrictions that would otherwise cause damage to the cell. Indeed, our current work Patteson, *et al*. [38] reveals that the loss of vimentin increases nuclear damage, such as nuclear membrane rupture, during migration through small pores. The effects of vimentin and cell stiffness on 3D migration and the relation between cell stiffness and migration speed indicates that vimentin is important for resisting compressive stresses and maintaining the structural integrity of the cell during migration. Vimentin's role in cell deformability and migration may therefore have wide ranging implications for nuclear damage and repair, genome expression, cell cycle, and cell fate, which are important to the maintenance of tissues and progression of diseases, including cancer. Overall, our findings demonstrate that vimentin's perinuclear stiffness controls 3d motility and provides new insight into how cells might alter their cytoskeleton to maximize invasion *in vivo* without compromising cell integrity.

**Methods**

*Cell Culture* Wild-type mouse embryo fibroblasts and vimentin-null mEF were kindly provided by J. Ericsson (Abo Akademi University, Turku, Finland) and maintained in DMEM with 25 mM HEPES and sodium pyruvate (Life Technologies; Grand Island, NY) supplemented by 10% fetal bovine serum, 1% penicillin streptomycin (Gibco), and nonessential amino acids (Life Technologies). For NIH-3T3 fibroblasts (American Type Culture Collection, Manassas, VA), cells were maintained in DMEM (Gibco) with 10% fetal bovine serum (ATCC) and 1% penicillin streptomycin, and 25 mM HEPES was added to media for micro-channel experiments. All cell cultures were maintained at 37°C and 5% $CO_2$.

*Immunofluorescence* Cells were fixed for immunofluorescence using 4% paraformaldehyde for 30 min at 37°C, permeabilized with 0.05% Triton X-100 solution in PBS (15 min., room temperature RT), and saturated with 1% serum albumin bovine (30 min, RT). For vimentin visualization, cells were incubated with primary anti-vimentin monoclonal rabbit antibody (1:200, RT, Abcam ab92547) or primary anti-vimentin polyclonal chicken antibody (1:200, RT, Novus NB300-223); secondary antibodies were anti-Rabbit Alexa Fluor 488 (1:1000, RT, Invitrogen A-11008) or anti-chicken Alexa Fluor 488 (1:1000, RT, Invitrogen A-11039). For visualizing microtubules, we use primary anti-tubulin monoclonal rat antibody (1:200, Serotec MCA77G) and secondary anti-rat AlexaFluor647 (Invitrogen A-21247). For immunostaining cells in micro-channels and on capillaries, primary antibodies were diluted to 1:1000 and kept overnight at 4°C. Cells were washed and stained with Rhodamine phalloidin 565 (Life Tech. r415) or Texas Red phalloidin and Hoechst 33342 (Molecular probes H-1399) or DAPI (Molecular probes) for 1 hr according to manufacturer's instructions. Cells were imaged with a Leica DMIRE2 inverted microscope with either a 40x (0.55 NA) air objective lens or 63x (0.70 NA) air objective lens.

*Microfluidic device fabrication and operation*
*Fabrication -* The microfluidic devices were built using standard soft lithography techniques and designed to prevent pressure gradients across the channels as described in Irimia, *et al.* [39]. The positive silicon master was generated by spinning KMPR 2010 onto silicon wafers (Wafer World Inc., West Palm Beach FL) to create a 10 μm thick layer. The photoresist was soft baked for 5 min at 95°C and exposed to UV light through a chrome mask (CAD/Art Services, Inc., **SI Fig. 2**) with a mask aligner (ABM-USE, Inc., ABM3000HR). Unexposed KMPR2010 was developed with SU-8 developer and rinsed with isopropanol. This process was repeated 2 times in order to stack the three layers of the device that were aligned with a mask aligner. The positive wafer obtained was silanized (Sigma 448931) in a vacuum chamber overnight. A PDMS (Sylgard 184) solution at a 1:10 ratio (curing:elastomer) was mixed and degassed. This solution was then poured over the silanized positive silicon wafer and baked 90 min at 80°C to generate a negative mold. The negative mold was silanized overnight in a vacuum chamber. This protocol was repeated to obtain the silanized positive mold that was used to build the microfluidic device. Finally, A PDMS mixture was degassed, poured over the PDMS positive molds, baked for 90 min at 80°C, and removed from the mold. The device was punched with 0.5 mm access holes for tubing inserts. Channels were sealed with a glass microscope slide using an oxygen plasma chamber (Femto-Diener Electronic).

*Operation-* To sterilize and clean the device, it was flushed with a solution of 70% (vol/vol) ethanol in deionized water ($diH_2O$), followed by rinsing with sterile $diH_2O$. Next, the device was submerged in sterile $diH_2Or$ and degassed to remove bubbles. Channels were then rinsed with phosphate buffered saline (PBS) and coated with surface ligands by pumping in a 50 ug/mL solution of either collagen 1 (BD Biosciences, Franklin Lakes, NJ) or fibronectin (purified from Salmon plasma) in PBS and incubating for 1 hr at 37°C. Finally, channels were then washed 3x with PBS, filled with cell culture media, and incubated at 37 °C for at least 30 min before seeding cells.

*Cell culturing and seeding-* Cells were trypsinized using 0.5% trypsin (GIBCO) at 37°C, centrifuged to remove trypsin, and re-suspended in cell media at densities of 10 million cells/mL. Using a hand-held syringe (Hamilton Company, 81320) and tubing (Hamilton Company, 90622), cells were gently pumped into the device inlet. Cells were preferentially placed near the opening of the channel constrictions by manually tilting the device for 2 to 4 min and allowing gravity to pull cells in suspension toward the constrictions (**SI Fig. 2&3**). Fluid reservoirs (barrel-less syringes (BD, Ref 309657) containing cell media) were connected to the channel outlets with tubing and arranged to ensure no pressure driven flow through the channels. The device was kept in a Tokai-Hit Imaging Chamber (Tokai Hit, Shizuoka-ken, Japan) and maintained at 37°C and 5% $CO_2$. Cells were allowed to adhere to channel surfaces for approximately 20-40 min before time-lapse imaging began. Time-lapse imaging was performed with a Leica DMIRE2 inverted microscope in bright-field with a 10x (0.3 NA) air lens. Images were taken every 4 min for 12-21 hr using an ASI *x/y/z* stage (MS – 2000, Applied Scientific Instrumentation) to capture multiple positions in the device.

### *Capillary fabrication and operation*
*Fabrication -* Capillaries having diameters ranging from 10-20 um were pulled from larger borosilicate glass capillaries with a diameter of 1.6 mm (Richland Glass, Richland, NJ) using a Narishige PB-7 pipette puller. Cell culture chambers (length, 60 mm, width 240mm, height, 12mm) were printed using either ABS or PLA plastic (Biomedical library, University of Pennsylvania). To allow for visualization of the cells, windows were designed into the upper and lower sections of the chambers. Capillaries were affixed to the inner surface of the chambers using UV-curable glue (NOA 68, Norland Products).

*Operation-* Capillaries were cleaned by rinsing once with ethanol, then rinsing with deionized $H_2O$, and then air dried. Chambers containing capillaries were then placed in a plasma cleaner (Harrick, PDC-32G) and exposed to air plasma for 5 mins. Overnight incubations with either 0.1 mg/ml collagen or fibronectin were then made while shaking on an orbital shaker. Capillaries were then rinsed 3X with PBS, sterilized under UV for 1 hr, and then incubated with culture media for at least 10 mins before cell seeding.

*Cell culturing and seeding -* Cells were trypsinized using 0.5% trypsin (GIBCO) at 37°C then centrifuged to remove trypsin. Resuspended cells at a density of 2.5 x $10^5$ cells/ml were added to the capillaries and allowed to attach for 60 mins at 37°C. Cell culture chambers were then transferred to a Tokai-Hit Imaging Chamber mounted on an ASI x/y/z stage and maintained at 37°C and 5% $CO_2$. Time-lapse images were taken at multiple positions every 10 mins for 18-21 hrs using a 40x objective.

### *Three-dimensional (3D) collagen gel preparation and imaging*
*Gel preparation -* Collagen gels (2 mg/mL) were prepared with final concentrations of 300 000 cells/mL by mixing together the following reagents in the order listed: pelleted and counted cells in media (10% v/v), 5x DMEM (20% v/v), FBS (10% v/v), 0.1 M NaOH (10% v/v), and 4 mg/mL collagen type 1 (Corning, REF 354236, 50 %v/v). Reagents are kept cold on ice while mixing. One mL of mixture was added to 20 mm dishes and maintained at 37°C and 5% $CO_2$. Cells were imaged in bright field, 24 hr after seeding in gel (Fig. 1).

*Imaging nucleus for tracking in gels* - To track cells in the 3D gels over time, the nuclei of vim +/+ and vim -/- mEF were fluorescently labeled with NLS-GFP. For these experiments, cells were transiently transfected with pEGF-C1-NLS. Forty-eight hr after transfection, cells were cultured in the collagen gel. Twenty-four hr later, cell nuclei were imaged at 10 min increments for 17 hr by using fluorescence microscopy and a 10x (0.3 NA) air lens.

### Cell area and motility measurements
*Cell area* - The projected cell area changed over time as the cell entered and moved through the channel (**SI Fig. 8**). Thus, we measured cell area at three designated points: (i) one hr before cells entered the channel, (ii) at the middle of the channel length, and (iii) one hr after cell exited the channel. Cell area was determined by manually tracing the periphery of single cells ($N$ = 25-35 cell per condition) using ImageJ software *(ImageJ Software, NIH, Bethesda, MD)*.

*Cell speed in micro-channels, glass slides, & capillaries* - Cell trajectories $r(x, y, t)$ were determined by tracking the center of mass of cells at either 4 or 10 min. increments using ImageJ Software (NIH) and the Manual Tracking plugin (https://imagej.nih.gov/ij/). Here, cell speed was determined over time $t$ as $|v(t)| = [r(t + \Delta t) - r(t)]/\Delta t$, where $\Delta t$ was approximately 30 min. To compare the directed cell motion in the 3D environments (micro-channels and capillaries) with the more random motion in 2D (glass slides), we chose the maximum value of $|v(t)|$ as the measure of cell speed for each cell in each experimental condition. For average cell speed and persistence in the micro-channels and glass slides, see **SI Fig. 5**. Cells that divided or moved out of the frame of view were excluded.

*Cell speed in collagen gels* – Because cell migration in the fibrillar collagen gels was limited, cell speed was determined as the cell displacement over $\Delta t$ = 4 hr, a time step large enough to measure displacements ($\approx$ 8 μm) but smaller than the cell persistence length: this method yielded approximately 50 measures of speed per cell by sampling over multiple times in a 17 hr video. Cells observed to move out of the field of view, divided, or died were excluded. Experiments were conducted three times for a total of 22-24 cells per condition.

**Cell stiffness measurements** Cell peri-nuclear stiffness measurements were performed using atomic force microscopy (NanoWizard 4, JPK) equipped with a liquid cell and temperature control setup. Silicon nitride cantilevers (ORC8, Bruker) with nominal 0.1 N/m spring constant and tip half-opening angle of 36° ± 2° were used for cell nanoindentation. Quantitative characterization of nanomechanical properties of the cells was realized by recording of multiple force vs distance curves in the peri-nuclear region (Fig. 4ii) with the constant force of 3 nN and indentation rate equal 0.4 Hz. Modified Hertz model was fitted to the data and Young`s modulus of each point was calculated as described previously [40].

**Actin and microtubule inhibitors** To determine the effects of microtubule and actin inhibitors on vim +/+ and vim -/- mEF migration in the micro-channels, the cells were treated with either the microtubule inhibitor nocodazole or actin inhibitor cytochalasin D (**SI Fig. 6 & 7**) while using time-lapse microscopy. Prior to administering inhibitors, cells were seeded in 10 μm collagen-coated micro-channels for three to six hr. Then, cells were treated with either 1 µg/mL nocodazole (Sigma) or 1 µg/mL cytochalasin D (Sigma) by diluting the reagents in cell media and gently pumping the solution through the micro-fluidic device with a hand-held syringe and tubing connected to the device inlet. Cell migration was observed for the following 17 hr, and the resultant average speed of the cells was determined for 8-13 cells per condition (1 experiment each). Control experiments with DMSO only were performed for each cell line. To ensure drugs did not diffuse out of the channel, a test using 40 nm fluorescent particles (Invitrogen F8770) as tracers was used to confirm stable concentrations during the experiment (**SI Fig. 7**). We found that nocodazole and cytochalasin D treatment decreases cell speed for both vim +/+ and vim -/- mEF in the micro-channels (**SI Fig. 6**).

*Transwell migration assays* Cells were seeded at sub-confluent concentrations (10-20 thousand cells/cm$^2$) on polycarbonate Transwell membranes with pore diameters of 3 µm (Corning, CLS3414), 5 µm (Corning, CLS3421), and 8 µm (Corning, CLS3428). Membranes were pre-coated with either collagen 1 (50 ug/mL) or fibronectin (50 ug/mL). After seeding cells, the filters were maintained at 37°C and 5% $CO_2$ for 15 hours. Cells were gently removed from either the top or bottom of the membrane with a cotton swab and immediately fixed with paraformaldehyde. To determine the number of cells per unit area, cells were stained with crystal violet and imaged at multiple locations across the membrane with a 10x objective. Cells were manually counted in 800x800 µm$^2$ fields of view (12-30 locations per condition). The percentage of cells that cross the membrane (Fig. 5) was then determined by taking the ratio of the number of cells on the bottom of the membrane and the sum of cells on the filter top and bottom.

*Estimate of compressional cell strain* – To estimate the cell strain through the Transwell filters, we assume that the cells maintain a constant volume of 1.76 pL (equivalent to a sphere of diameter 15 µm). This volume is less than the value determined by multiplying the average cell spread area by the channel height, which yields an overestimate of 2.1 +/- 0.2 pL for vim +/+ mEF and 2.4 +/- 0.2 pL for vim -/- mEF (**SI Fig. 8**). Compressional cell strain is then estimated as the magnitude of $(L_1-L)/L$, where $L$ is the cell size in the unstressed spherical state (15 µm) and $L_1$ is the narrowest dimension of the cell while crossing the pore, the diameter of the pore (*c.f.* Fig. 5).

***Statistical Analysis*** Data presented as mean values ± standard errors (SE). Each experiment was performed a minimum of two times unless otherwise stated. The unpaired Student's *t*-test with two tails at the 95% confidence interval was used to determine statistical significance. Denotations: *, $p <= 0.05$; **, $p < 0.01$; ***, $p < 0.001$; ns, $p > 0.05$. The Fisher's exact test was used to confirm statistical significance between the proportion of cells that exit the channels or exhibit twirling behavior (Fig. 2c).

**SUPPLEMENTARY MOVIES**

**SI Movie 1: Spontaneous migration of wild-type mEF crossing micro-channel.**
Wild-type mEF migrating through a 10x10 µm$^2$ micro-fluidic constriction, coated with collagen-I. The length of the video is 9.3 hr. Images are taken at 4-minute increments with a 10x air objective.

**SI Movie 2: Spontaneous migration of vimentin-null mEF crossing micro-channel.**
Vimentin-null mEF migrating through a 10x10 μm$^2$ micro-fluidic constriction, coated with collagen-I. The length of the video is 6.3 hr. Images are taken at 4-minute increments with a 10x air objective.

**SI Movie 3: Wild-type mEF that does not cross channel.**
Wild-type mEF interacting with a micro-fluidic constriction. The cell does not enter the channel, but instead switches directions. Channel coated with collagen-I. The length of the video is 9.2 hr. Images are taken at 4-minute increments with a 10x air objective.

**SI Movie 4: Twirling motions of wild-type mEF in micro-channel.**
Wild-type mEF migrating through a 10x10 μm$^2$ micro-fluidic constriction, coated with fibronectin. The cell appears to twirl around the channel, moving in a three-dimensional spiral. The length of the video is 6.1 hr. Images are taken at 4-minute increments with a 10x air objective.

**Acknowledgements** This work was supported by the National Institutes of Health-National Institute of General Medical Sciences (P01 GM096971) and National Science Center, Poland under Grant: UMO-2015/17/B/NZ6/03473. We thank Robert Goldman for kindly sharing the pEGF-C1-NLS plasmid construct and Dennis Discher and Manu Tewari for help amplifying the plasmid. We also thank Eric Johnston and Nathan Bade for help developing the micro-fluidic device and Mateusz Cieśluk for his technical assistance during AFM experiments.

**Author Contributions** A.E.P. designed, performed, and analyzed motility experiments using micro-fluidics devices, collagen gels, and Transwell membrane assays. F.B. designed and performed capillary experiments and analysis. K.P., R.B., and P.D designed and performed AFM measurements and analysis and K.P. assisted in collagen gel preparation. E.C. assisted with cell culture, and E.C. and P.G. developed micro-fluidic device. A.E.P., F.B., K.P., E.C., P.G., and P.A.J. contributed to project design and wrote the manuscript.